\documentclass[aip,apl,reprint,noshowpacs,noshowkeys,amssymb,amsmath,amsfonts,bibnotes]{revtex4-2}

\usepackage{graphicx}
\usepackage{theorem}
\usepackage{dcolumn}
\usepackage{longtable}
\usepackage{bm}
\usepackage{accents}
\usepackage{color}
\usepackage[dvipsnames]{xcolor}
\usepackage{float}
\usepackage{booktabs}

\usepackage{siunitx}
\newcommand{\SSS}{\scriptscriptstyle}

\newcommand{\etal}{{\textit{et al.}}}

\newcommand{\kB}{k_{\SSS\text{B}}}

\renewcommand{~}{\,}

\makeatletter
\renewcommand{\figurename}{Fig.}
\renewcommand{\tablename}{Table}
\renewcommand{\fnum@figure}[1]{\textbf{\figurename~\thefigure.} }
\renewcommand{\fnum@table}[1]{\textbf{\tablename~\thetable.} }
\makeatother

\begin{document}

\title{Telecom-Band Optical Transitions of Erbium Atoms in Rare-Gas Solids}

\author{Evgenii Zaitsev}
\affiliation{Department of Physics and Astronomy, University of Notre Dame, Notre Dame, Indiana 46556, USA\looseness=-1}

\author{Alec Dinerstein}
\affiliation{Department of Physics and Astronomy, University of Notre Dame, Notre Dame, Indiana 46556, USA\looseness=-1}

\author{Shan Zou}
\affiliation{Department of Physics and Astronomy, University of Notre Dame, Notre Dame, Indiana 46556, USA\looseness=-1}

\author{Charles Peretti}
\affiliation{National High Magnetic Field Laboratory, 1800 East Paul Dirac Drive, Tallahassee, Florida 32310, USA\looseness=-1}
\affiliation{Department of Mechanical and Aerospace Engineering, FAMU-FSU College of Engineering, Florida State University, Tallahassee, Florida 32310, USA\looseness=-1}
\affiliation{FSU Quantum Initiative, Florida State University, Tallahassee, FL 32310, USA\looseness=-1}

\author{Yutian Wen}
\affiliation{Department of Physics and Astronomy, University of Notre Dame, Notre Dame, Indiana 46556, USA\looseness=-1}

\author{Wei Guo}
\affiliation{National High Magnetic Field Laboratory, 1800 East Paul Dirac Drive, Tallahassee, Florida 32310, USA\looseness=-1}
\affiliation{Department of Mechanical and Aerospace Engineering, FAMU-FSU College of Engineering, Florida State University, Tallahassee, Florida 32310, USA\looseness=-1}
\affiliation{FSU Quantum Initiative, Florida State University, Tallahassee, FL 32310, USA\looseness=-1}

\author{Gregory V. Hartland}
\affiliation{Department of Chemistry and Biochemistry, University of Notre Dame, Notre Dame, Indiana 46556, USA\looseness=-1}

\author{Yizhong Huang}\email[Email: ]{huangy@anl.gov}
\affiliation{Center for Nanoscale Materials, Argonne National Laboratory, Lemont, Illinois 60439, USA\looseness=-1}

\author{Dafei Jin}\email[Email: ]{dfjin@nd.edu}
\affiliation{Department of Physics and Astronomy, University of Notre Dame, Notre Dame, Indiana 46556, USA\looseness=-1}

\date{\today}

\begin{abstract}

    Rare-earth (RE) atoms, such as erbium (Er), have inner-shell optical transitions in the near-infrared, some of which are in the telecom bands. These transitions are largely insensitive to the host environment due to effective shielding by the filled outer shells. Cryogenic rare-gas (RG) solids, such as solid neon (Ne) and argon (Ar), combine high chemical purity with a nearly nuclear-spin-free environment, making them attractive solid hosts for preserving optical and spin coherence. In this work, we isolate neutral Er atoms in solid Ne and Ar at \qty{6}{\K} and characterize their emission by steady-state and time-resolved photoluminescence (PL) spectroscopy. In both hosts, we observe several near-infrared transitions between \qtyrange{1000}{1500}{\nm}, including inner-shell emission in the telecom O and E bands. The $4f \rightarrow 5d$ transition near \qty{1299}{\nm} forms a multiplet of narrow components, consistent with crystal-field splitting, and decays on a microsecond timescale. The $4f \rightarrow 4f$ transition near \qty{1437}{\nm} shows a matrix shift below \qty{1}{\per\cm} in both hosts. In solid Ne, its inhomogeneous linewidth is bounded by the spectrometer resolution to \qty{<22}{\GHz}, and its PL lifetime reaches up to \qty{78.9}{\ms}, close to the estimated radiative limit. Thermal annealing extends the \qty{1437}{\nm} lifetime in Ne and suppresses the phonon sidebands relative to the zero-phonon lines in Ar. These results identify RG solids as promising hosts for long-lived telecom-band Er emission and provide a foundation for future spin--photon interfaces and hybrid quantum architectures.

\end{abstract}

\maketitle
\pretolerance=9000

\subsection{Introduction}

Solid-state spin--photon interfaces are fundamental building blocks for quantum networks, as they coherently link long-lived stationary qubits to optical photons for entanglement distribution and quantum communication. Several promising platforms have been developed, including color centers in diamond~\cite{jelezko2006single, togan2010quantum, pompili2021realization, li2024heterogeneous} and silicon carbide~\cite{koehl2011room, anderson2019electrical, widmann2015coherent, lukin20204h}, T centers in silicon~\cite{higginbottom2022optical, song2026entanglement}, semiconductor quantum dots~\cite{gao2012observation}, and rare-earth ions in solids~\cite{saglamyurek2011broadband, ortu2018simultaneous, thiel2011rare, zhong2017nanophotonic, ruskuc2022nuclear}. Among these, rare-earth ions such as $\mathrm{Yb}^{3+}$, $\mathrm{Tm}^{3+}$, and $\mathrm{Er}^{3+}$ are particularly attractive. Their optically active $4f$ electrons are shielded by the filled $5s^2$ and $5p^6$ subshells, so their optical transitions are largely insensitive to the host environment.

The $4f \rightarrow 4f$ transitions, in particular, keep their characteristic wavelengths across a wide range of crystalline hosts. Many of them lie in the near-infrared (NIR), including wavelengths compatible with fiber-optic telecommunications. For example, $\mathrm{Er}^{3+}$ ions in solid hosts emit in the telecom C band near \qty{1.53}{\um}. They have demonstrated electron spin coherence times of tens of milliseconds~\cite{le2021twenty}, nuclear spin coherence times exceeding one second~\cite{ranvcic2018coherence}, and optical homogeneous linewidths below \qty{100}{\hertz}~\cite{bottger2009effects}. Individual $\mathrm{Er}^{3+}$ ions have also been detected and read out through integration with nanophotonic cavities~\cite{dibos2018atomic, raha2020optical}, single-electron transistors~\cite{yin2013optical}, and single-microwave-photon detectors~\cite{wang2023single}. This combination of telecom-band emission, long spin coherence, and compatibility with solid-state devices has made $\mathrm{Er}^{3+}$ a versatile quantum platform, enabling indistinguishable telecom-band single photon sources~\cite{ourari2023indistinguishable}, spin qubits~\cite{chen2020parallel, gupta2025dual, gritsch2025optical}, coherent nuclear spin control~\cite{uysal2023coherent}, and microwave-to-optical quantum transduction~\cite{rochman2023microwave}.

Despite this progress, the host material remains a major source of decoherence and inhomogeneous broadening~\cite{awschalom2018quantum, wolfowicz2021quantum, gupta2025dual}. Charge noise, magnetic impurities, nuclear spins of the host lattice, and fabrication-induced disorder can broaden optical linewidths and degrade spin coherence. An ideal host should therefore combine chemical and magnetic purity with minimal perturbation of the quantum emitter, while providing optical transparency, cryogenic compatibility, and seamless integration with photonic and microwave devices.

Rare gases such as Ne and Ar form van der Waals crystals below their triple-point temperatures of \qty{24.56}{\K} and \qty{83.81}{\K}, respectively. Their closed-shell electronic configurations make these solids chemically inert and only weakly polarizable, and they can be grown with very low impurity concentrations. They also have an almost nuclear-spin-free lattice. Of the three stable Ne isotopes, only $^{21}$Ne has a nonzero nuclear spin ($I=3/2$), with a natural abundance of only 0.27\%. All three stable Ar isotopes, $^{36}$Ar, $^{38}$Ar, and $^{40}$Ar, have zero nuclear spin. These intrinsic properties render solid Ne and Ar ultra-low-noise host matrices, making them promising for preserving quantum coherence. In our previous studies, single electrons trapped on the surface of solid Ne showed charge-qubit coherence times of about \qty{0.1}{\ms}, with theoretically estimated spin coherence times exceeding \qty{1}{\s}~\cite{zhou2022single, Zhou2024, li2026solid, chen2022electron}. 

Alkali atoms such as sodium (Na), rubidium (Rb), and cesium (Cs) embedded in solid Ne can retain long spin coherence times close to those of free atoms~\cite{PhysRevResearch.6.L012048, PhysRevA.104.032611, PhysRevJourArch.10.1103, upadhyay2020ultralong, kanagin2025impurities}. However, their optical transitions involve unshielded outer-shell electrons, which couple strongly to the lattice and cause severe spectral broadening.

RE atoms can overcome this limitation because their inner-shell transitions are shielded. For instance, Yb atoms in solid Ne at \qty{4}{\K} show long multi-exponential lifetimes of \num{17}, \num{67}, and \qty{315}{\s}, because local Stark mixing enhances population transfer into the metastable $6s6p\,{}^3\mathrm{P}_0$ state by seven orders of magnitude relative to free atoms~\cite{PhysRevLett.107.093001, xu2014measurement}. Tm atoms isolated in solid He, Ne, and Ar exhibit optical lifetimes exceeding \qty{10}{\ms} for the \qty{1140}{\nm} $4f \rightarrow 4f$ magnetic-dipole transition \cite{PhysRevA.99.022505, PhysRevB.56.780}. Among RE atoms, Er is especially compelling. Its large magnetic dipole moment, rich electronic structure, and narrow optical transitions enable a novel platform for quantum simulation and quantum computing, as demonstrated with optical tweezer arrays of Er atoms \cite{grun2024optical, PhysRevResearch.3.033256, su2023dipolar, su2026topological}. Importantly, inner-shell transitions of neutral Er fall directly in the telecom O and E bands, offering a natural interface for fiber-based quantum networks. To our knowledge, Er atoms in rare-gas matrices have not yet been studied.

Here, we demonstrate matrix-isolated neutral Er atoms in solid Ne and Ar as a promising telecom quantum material platform. Using PL spectroscopy at a sample temperature of about \qty{6}{\K}, we characterize their emission from the visible (VIS) to the NIR range. We resolve several narrow inner-shell transitions between \num{1000} and \qty{1500}{\nm}, including emission in the telecom O and E bands. The measured excited-state lifetimes range from microseconds to tens of milliseconds and depend on the matrix preparation and thermal annealing. In particular, the $4f \rightarrow 4f$ transition near \qty{1437}{\nm} lies in a region where standard optical fibers have a low loss of about \qty{0.3}{\dB\per\km}. In solid Ne, this transition shows a minimal matrix shift, a resolution-limited inhomogeneous linewidth \qty{<22}{\GHz} (\qty{\approx0.15}{\nm}), and a lifetime of up to \qty{79}{\ms}. These results identify RG solids as promising hosts for narrow telecom emission from Er. They provide a spectroscopic foundation for future studies of optical coherence, spin properties, and integration with photonic and superconducting quantum architectures.

\subsection{Experimental setup}

\begin{figure*}
    \centering
    \includegraphics[width=\textwidth]{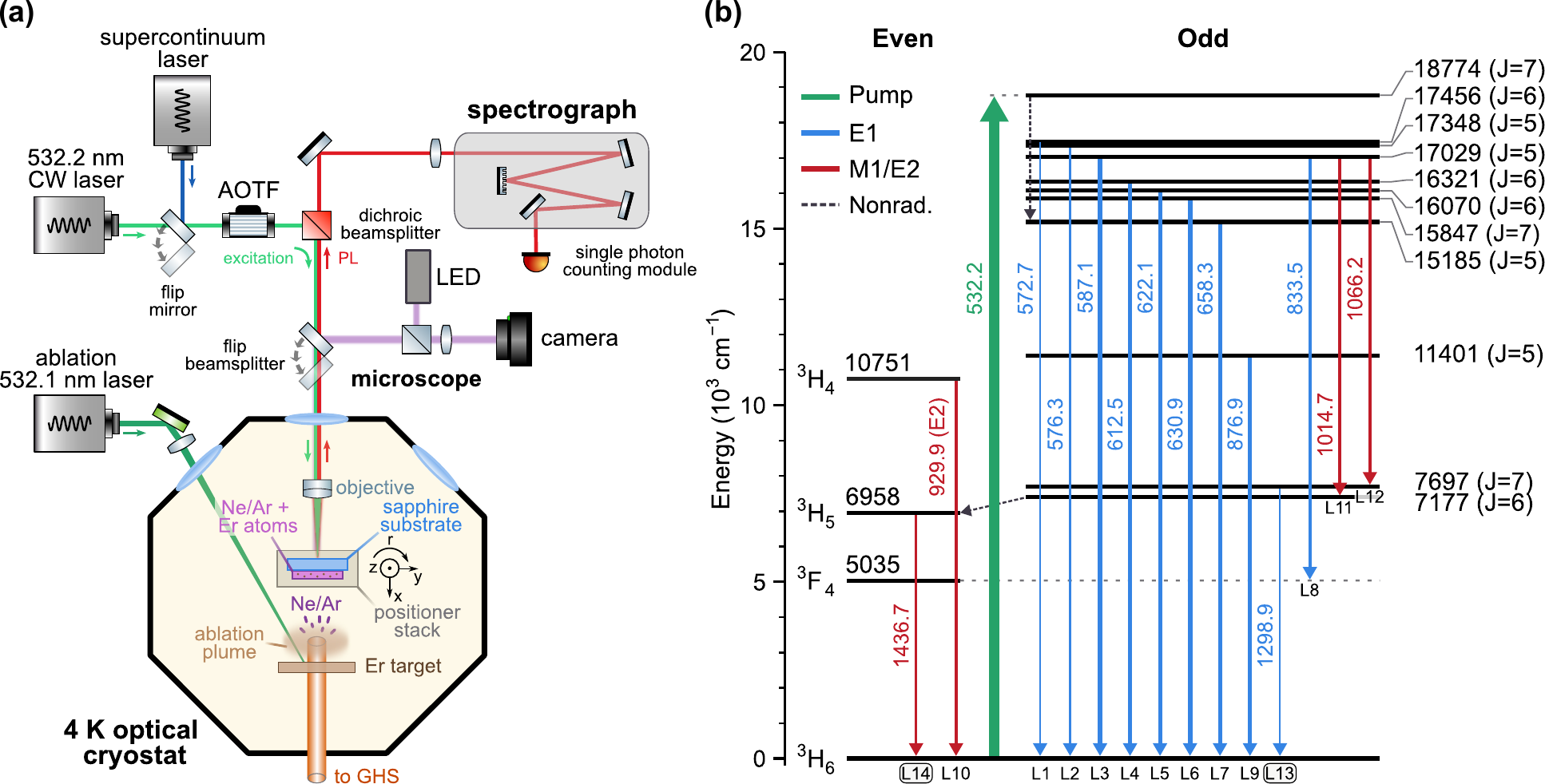}
    \caption{\textbf{(a)} Experimental setup. Er atoms produced by laser ablation are co-deposited with Ne or Ar onto a cold sapphire substrate. PL is excited by a \qty{532.2}{\nm} CW laser or an AOTF-filtered supercontinuum source and detected with a spectrograph and single-photon detectors. \textbf{(b)} Simplified energy-level diagram of neutral Er~\cite{atomic_energy_levels}, with even- and odd-parity levels in separate columns and energies in \unit{\per\cm}. The green arrow marks the \qty{532.2}{\nm} excitation. Blue and red arrows show the observed E1 and M1/E2 transitions L1--L14, respectively, labeled with their free-atom air wavelengths in nm; the boxed labels mark the telecom-band transitions L13 and L14. Arrow thickness indicates the relative PL intensity under \qty{532.2}{\nm} excitation. Dashed arrows indicate nonradiative relaxation through the dense ladder of odd-parity levels below the pumped level at \qty{18774}{\per\cm}.}
    \label{Fig:Fig1}
\end{figure*}

Fig.~\ref{Fig:Fig1}(a) shows the experimental setup. A tabletop closed-cycle optical cryostat reaches a base temperature of \qty{4}{\K} and provides optical access through three windows. Inside the cryostat, a single-crystal sapphire substrate is mounted on a cryogenic positioner stack that allows three-axis translation and \ang{360} rotation. The heat load of the positioner stack raises the sample temperature to \qtyrange{5}{6}{\K}. The rare gas is delivered to the substrate through a custom gas-handling system, whose geometry and thermal anchoring are optimized for growing solid Ne and Ar. During matrix growth, Er atoms are produced by pulsed laser ablation of a high-purity (\qty{99.99}{\percent}) Er foil. For ablation, we use a focused, frequency-doubled $Q$-switched Nd:YAG laser at \qty{532}{\nm}, with a pulse duration of \qty{5}{\ns} and a pulse energy of about \qty{25}{\milli\joule}. The ablated Er atoms and the rare gas are co-deposited onto the cold substrate to form matrix-isolated Er samples.

Steady-state PL, PL excitation (PLE), and time-resolved PL measurements share a common excitation and collection path. A single-longitudinal-mode Nd:YVO$_4$ laser provides CW excitation at \qty{532.2}{\nm}. For wavelength-selective excitation between \qtyrange{450}{700}{\nm}, we use a supercontinuum laser filtered by an AOTF with a passband of about \qty{2}{\nm}. The AOTF also gates both excitation sources for PL decay measurements, with a fall time (90--10\%) of about \qty{1.2}{\us}. Inside the vacuum chamber, an achromatic doublet focuses the excitation to a spot of about \qty{0.1}{\mm} on the sample and collects the PL. The PL is dispersed by a \qty{500}{\mm} spectrograph with a \qty{1200}{grooves\per\mm} grating and detected by room-temperature single-photon detectors, a Si detector for the VIS range and an InGaAs detector for the NIR range. Depending on the slit width, the spectral resolution is about \qtyrange{0.1}{0.2}{\nm}.

\subsection{Spectroscopic Measurements}

\begin{figure*}[t]
    \centering
    \includegraphics[width=\textwidth]{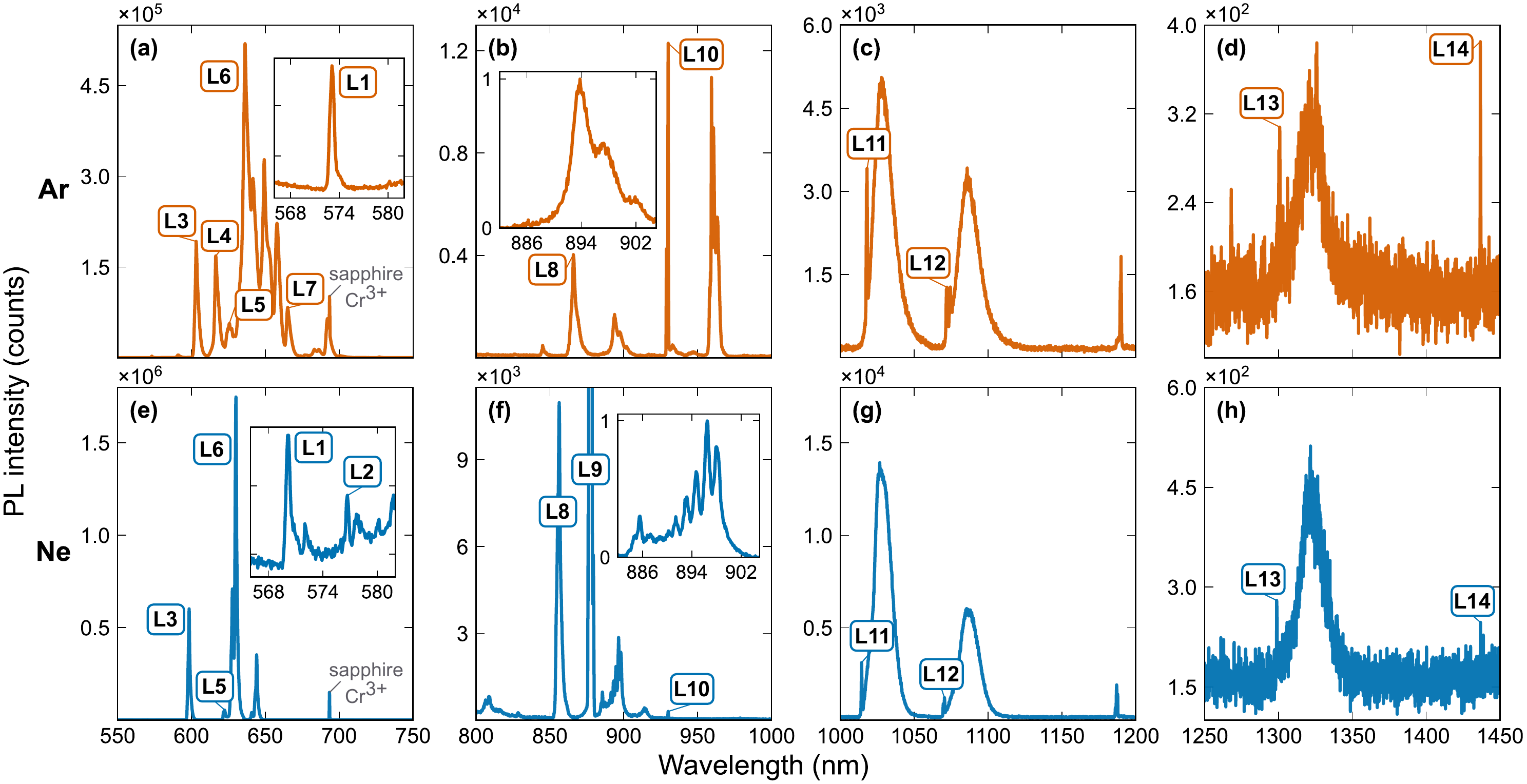}
    \caption{PL spectra of Er in solid Ar \textbf{(a--d)} and Ne \textbf{(e--h)} at \qty{6}{\K} under \qty{532.2}{\nm} CW excitation, covering the \qtyrange{550}{750}{\nm}, \qtyrange{800}{1000}{\nm}, \qtyrange{1000}{1200}{\nm}, and \qtyrange{1250}{1450}{\nm} ranges. Labels L1--L14 mark the assigned transitions shown in Fig.~\ref{Fig:Fig1}(b); unlabeled features could not be reliably assigned to known free-atom transitions. Insets enlarge the weak features near \qtyrange{566}{582}{\nm} and compare the structure near \qtyrange{882}{905}{\nm} in the two hosts; the latter insets are baseline-subtracted and peak-normalized.}
    \label{Fig:Fig2}
\end{figure*}

Steady-state PL spectroscopy probes the electronic transitions of matrix-isolated Er atoms and their sensitivity to the local RG environment. Fig.~\ref{Fig:Fig1}(b) summarizes the free-atom levels relevant to the main spectral features, with the observed transitions labeled L1--L14 as in Fig.~\ref{Fig:Fig2}. Neutral Er has an even-parity $4f^{12}6s^2$ ground configuration with a ${}^3\mathrm{H}_6$ ground level ($J=6$)~\cite{atomic_energy_levels}. The \qty{532.2}{\nm} excitation is blue-detuned by about \qty{11}{\per\cm} from the E1-allowed transition to the odd-parity $4f^{11}5d6s^2\,(11/2,3/2)^\circ_7$ level at \qty{18774.123}{\per\cm}, whose free-atom air wavelength is \qty{532.500}{\nm}. This detuning is small compared with typical matrix shifts and broadening, so the laser likely excites this transition directly. The PLE spectra of L8 and L11--L14 support this excitation pathway: all show a strong enhancement between \qtyrange{530}{555}{\nm} (not presented in this paper). 

Fig.~\ref{Fig:Fig2} shows representative PL spectra of Er in solid Ar and Ne, recorded at about \qty{6}{\K} under \qty{10}{\milli\watt} CW excitation at \qty{532.2}{\nm}. The sharp line near \qty{693.5}{\nm}, present in both hosts, is the $R$-line emission of trace $\mathrm{Cr}^{3+}$ impurities in the sapphire substrate~\cite{Zykova2022Photoluminescence}. The remaining features depend on the host and on the growth conditions. Compared with the free-atom transitions, the features in Ar are typically broader and red-shifted, whereas those in Ne are sharper and mostly lie close to the free-atom positions. This contrast is consistent with a stronger host perturbation in Ar, whose static dipole polarizability ($\alpha_{\mathrm{Ar}}=\qty{1.64}{\angstrom\cubed}$) is about four times that of Ne ($\alpha_{\mathrm{Ne}}=\qty{0.40}{\angstrom\cubed}$)~\cite{Derevianko2010Polarizabilities,Schwerdtfeger2019Polarizabilities}.

In both hosts, the strongest visible features are L3 and L6, followed by L5 in Ne and by L4, L5, and L7 in Ar. In contrast, L1 and L2 are about three orders of magnitude weaker than L6. This ordering largely follows the energy gap below each upper level rather than the free-atom line strength. Below the pumped level at \qty{18774}{\per\cm}, the odd-parity levels form a dense ladder, and the population relaxes nonradiatively toward lower levels. The rate of this multiphonon relaxation drops steeply with the number of phonons needed to bridge the gap to the next lower level~\cite{Riseberg1968}. Because the maximum phonon energies of solid Ne and Ar are only about \qtyrange{50}{65}{\per\cm}~\cite{Endoh1975NePhonons,Fujii1974ArPhonons}, a level emits efficiently only if a large gap lies below it. The upper levels of L1 and L2 lie only \qty{108}{\per\cm} and \qty{50}{\per\cm} above their nearest lower neighbors, so they are strongly quenched. By contrast, the upper levels of L5 and L3 lie \qty{224}{\per\cm} and \qty{302}{\per\cm} above the next level. The upper level of L6 lies \qty{662}{\per\cm} above it, the largest gap in this part of the ladder, and L6 is correspondingly the strongest line in both hosts. For convenience, we refer to this ladder as the upper manifold and to the levels below \qty{11500}{\per\cm} as the lower manifold.

The same argument explains the behavior of the lower manifold. The lowest level of the upper manifold lies about \qty{2000}{\per\cm} above the next lower level (not shown in Fig.~\ref{Fig:Fig1}(b)), a gap that multiphonon relaxation cannot bridge, so the lower manifold must be populated radiatively. The observed L11 and L12 transitions provide one such pathway, terminating on the odd levels at \qty{7177}{\per\cm} and \qty{7697}{\per\cm}. E1 branches from the upper levels to ${}^3\mathrm{H}_5$ and ${}^3\mathrm{F}_4$, such as L8, feed the even levels directly. The strong dependence of L9 on excitation intensity suggests that its upper level at \qty{11401}{\per\cm} is populated mainly by excited-state absorption. Within the lower manifold, the upper level of L13 (\qty{1298.9}{\nm}) lies \qty{520}{\per\cm} above the odd $J=6$ level at \qty{7177}{\per\cm}. Its emission is therefore observable but strongly quenched, consistent with the microsecond-scale decay discussed in Sec.~\ref{Sec:Lifetime}. The $J=6$ level itself lies only \qty{218}{\per\cm} above ${}^3\mathrm{H}_5$ and relaxes before it can emit. This explains why its E1 transition to the ground state at \qty{1393.1}{\nm} is not observed. Finally, ${}^3\mathrm{H}_5$ lies \qty{1923}{\per\cm} above ${}^3\mathrm{F}_4$, a gap that would require about 30--40 phonons. Nonradiative decay from this level is therefore negligible, and population accumulates there. ${}^3\mathrm{H}_5$ thus acts as the main bottleneck of the lower manifold and gives rise to the long-lived L14 emission at \qty{1436.7}{\nm}.

The emission near \qty{890}{\nm} shows pronounced structure in solid Ne (Fig.~\ref{Fig:Fig2}(f), inset), consisting of two overlapping series. The first starts at \qty{885.6}{\nm} and continues through three weaker components. Its spacing decreases from about \qty{24}{\per\cm} to \qty{15}{\per\cm}, and its intensity falls away from the first component. The second series comprises five components between \qty{891.4}{\nm} and \qty{898.0}{\nm} with a nearly constant spacing of \qty{20.6}{\per\cm} (\qty{2.6}{\meV}). Its intensity rises to a maximum at the fourth component (\qty{896.5}{\nm}) and then falls. The break in spacing near \qty{891}{\nm} is reproduced in independent samples, while the relative weight of the two series varies between samples. The two series also have different excitation spectra (not presented in this paper). Emission near \qtyrange{882}{886}{\nm} is excited most efficiently at about \qty{570}{\nm} and \qty{660}{\nm}, whereas emission near \qtyrange{893}{899}{\nm} is excited at about \qty{550}{\nm}, \qty{640}{\nm}, and \qty{670}{\nm}. We therefore attribute the structure to emitters in two distinct trapping sites, each coupled to a pseudolocalized vibrational mode of the Er impurity. Both mode energies lie well within the phonon band of solid Ne~\cite{Endoh1975NePhonons}, and resolved phonon sidebands of Yb atoms in solid Ne provide a precedent for such modes~\cite{PhysRevA.104.062809}. The electronic origins of the two series remain unassigned.

In solid Ar, the emission in this region is dominated by a broad maximum at \qty{893.8}{\nm}, followed on the red side by a weaker resolved maximum near \qty{897.2}{\nm} and a shoulder near \qty{901.5}{\nm} (Fig.~\ref{Fig:Fig2}(b), inset). Their spacing of about \qtyrange{40}{55}{\per\cm} lies within the phonon band of solid Ar~\cite{Fujii1974ArPhonons}, so these features may be phonon replicas of a local mode, analogous to the progressions in Ne. Compared with Ne, however, the components are broader and much less distinct, consistent with stronger electron--phonon coupling and faster dephasing in Ar. The spectra cannot exclude contributions from inhomogeneous broadening or from overlapping electronic and site components.

Two broad bands appear near \qty{1028}{\nm} and \qty{1086}{\nm}, each with a narrow feature on its short-wavelength side. We attribute the broad bands to phonon sidebands and assign the narrow features to their zero-phonon lines, L11 and L12. These are two branches from the odd-parity $4f^{11}5d6s^2$ level at \qty{17029.058}{\per\cm} ($J=5$) to the odd-parity levels at \qty{7176.503}{\per\cm} ($J=6$) and \qty{7696.956}{\per\cm} ($J=7$)~\cite{atomic_energy_levels}. Both branches connect levels of the same parity and are therefore E1-forbidden in the free atom. The $J=5\rightarrow6$ branch is M1-allowed, whereas the $J=5\rightarrow7$ branch requires an E2 contribution or matrix-induced mixing. The free-atom energies give air wavelengths of \qty{1014.687}{\nm} and \qty{1071.276}{\nm}. The measured origins in Ne lie at \qty{1014.4}{\nm} and \qty{1070.5}{\nm}, corresponding to small matrix shifts of about $+\qty{2.8}{\per\cm}$ and $+\qty{6.8}{\per\cm}$. Their separation of about \qty{516.5}{\per\cm} is close to the free-atom lower-level separation of \qty{520.453}{\per\cm}. The small shifts are consistent with all three levels belonging to the same configuration, so that their matrix shifts largely cancel. PLE measurements further support the assignment (not presented in this paper). The two bands have identical excitation spectra, indicating a common upper level. They are excited through Er ground-state absorption lines, including direct excitation near \qty{585}{\nm} on the L3 transition to the \qty{17029}{\per\cm} level. The excitation efficiency drops to the background between about \qty{588}{\nm} and \qty{600}{\nm}, which places the emitting level near \qty{17000}{\per\cm}.

A prominent emission group near \qty{1299}{\nm} (L13), in the telecom O band, arises from decay of the predominantly $4f^{11}5d6s^2$ odd-parity $J=7$ level at \qty{7697}{\per\cm} to the $4f^{12}6s^2\,{}^3\mathrm{H}_6$ ground level. The corresponding free-atom air wavelength is \qty{1298.860}{\nm}~\cite{atomic_energy_levels}. Patscheider \etal\ directly observed and coherently drove this inner-shell $4f\rightarrow5d$ transition in ultracold Er atoms. They measured an excited-state lifetime of \qty{178(19)}{\ms}, corresponding to a natural linewidth of \qty{0.9(1)}{\hertz}~\cite{PhysRevResearch.3.033256}. These free-atom values provide a reference for the linewidths and lifetimes we observe in the matrices.

Table~\ref{tab:1299_components} summarizes the main spectral features in both hosts. In Ne, eight features span about \qty{23}{\per\cm}. The strongest component (Ne5) lies within \qty{0.4}{\per\cm} of the free-atom transition, and the multiplet as a whole is centered on it. In Ar, the structure is more strongly blended, spreads over about \qty{31}{\per\cm}, and is shifted to the red by about \qty{10}{\per\cm}. Several effects may contribute to this structure, including crystal-field splitting of the upper and lower levels and inequivalent trapping sites. A comparison with L14 at \qty{1437}{\nm}, discussed below, helps separate these contributions. L14 also terminates on the ${}^3\mathrm{H}_6$ ground level but shows no comparably extended multiplet. The structure of L13 therefore originates mainly from its upper level. This is expected, because the $5d$ electron of the $4f^{11}5d6s^2$ upper level interacts more strongly with the matrix than the shielded $4f$ electrons do. This leads to a larger crystal-field splitting and larger site-dependent shifts.

\begin{table}[tb]
\centering
\caption{Main spectral features of the L13 emission group. The shift $\Delta\tilde{\nu}=\tilde{\nu}-\tilde{\nu}_{0}$ is given relative to the free-atom transition at $\tilde{\nu}_{0}=\qty{7696.956}{\per\cm}$ (\qty{1298.860}{\nm} in air); positive values correspond to higher photon energy. Bold entries mark the strongest component in each host, used for the lifetime measurements. Rows are aligned for compactness only and do not imply a correspondence between the hosts.}
\label{tab:1299_components}
\begin{tabular*}{\columnwidth}{@{\extracolsep{\fill}}lcc lcc@{}}
\toprule
\multicolumn{3}{c}{\textbf{Ar}} & \multicolumn{3}{c}{\textbf{Ne}} \\
\cmidrule(lr){1-3}\cmidrule(lr){4-6}
ID & $\lambda_{\mathrm{meas}}$ & $\Delta\tilde{\nu}$ & ID & $\lambda_{\mathrm{meas}}$ & $\Delta\tilde{\nu}$ \\
   & (\unit{\nm}) & (\unit{\per\cm}) & & (\unit{\nm}) & (\unit{\per\cm}) \\
\midrule
Ar1  & 1298.28 & $+3.4$  & Ne1 & 1296.95 & $+11.3$ \\
Ar2  & 1298.51 & $+2.1$  & Ne2 & 1297.44 & $+8.4$  \\
Ar3  & 1299.10  & $-1.4$  & Ne3 & 1298.06 & $+4.7$  \\
Ar4  & 1299.41 & $-3.3$  & Ne4 & 1298.44 & $+2.5$  \\
\textbf{Ar5} & \textbf{1300.19} & $\bm{-7.9}$ & \textbf{Ne5} & \textbf{1298.79} & $\bm{+0.4}$ \\
Ar6  & 1300.88 & $-12.0$ & Ne6 & 1299.12 & $-1.5$  \\
Ar7  & 1302.00 & $-18.6$ & Ne7 & 1299.82 & $-5.7$  \\
Ar8  & 1302.61 & $-22.2$ & Ne8 & 1300.85 & $-11.8$ \\
Ar9  & 1302.91 & $-23.9$ &     &         &         \\
Ar10 & 1303.54 & $-27.6$ &     &         &         \\
\bottomrule
\end{tabular*}
\end{table}

Broad emission also appears on the red side of L13, with maxima near \qty{1318.6}{\nm} in Ne and \qty{1325.5}{\nm} in Ar. These maxima lie about \qty{120}{\per\cm} and \qty{150}{\per\cm} below the strongest L13 components, roughly twice the one-phonon cutoff of either host. We tentatively attribute them to multiphonon sidebands, although emission from other sites or transitions may also contribute.

L14 near \qty{1437}{\nm} behaves very differently from L13. We assign it to the $4f^{12}6s^2\,{}^3\mathrm{H}_5\rightarrow{}^3\mathrm{H}_6$ transition, whose free-atom energy of \qty{6958.329}{\per\cm} corresponds to an air wavelength of \qty{1436.734}{\nm}~\cite{atomic_energy_levels}. The emission maximum lies at \qty{1436.74}{\nm} in Ne and \qty{1436.70}{\nm} in Ar. These positions correspond to shifts of only about $-\qty{0.03}{\per\cm}$ and $+\qty{0.16}{\per\cm}$, both within the calibration accuracy of our spectrometer. This close agreement indicates weak host perturbation of the $4f^{12}$ configuration, whose optically active electrons are shielded by the filled $5s^2$ and $5p^6$ subshells. Apart from the closely spaced components discussed below, we detected no other narrow or broad components attributable to this transition over the wider scanned range.

\begin{figure}[htbp]
    \centering
    \includegraphics[width=\columnwidth]{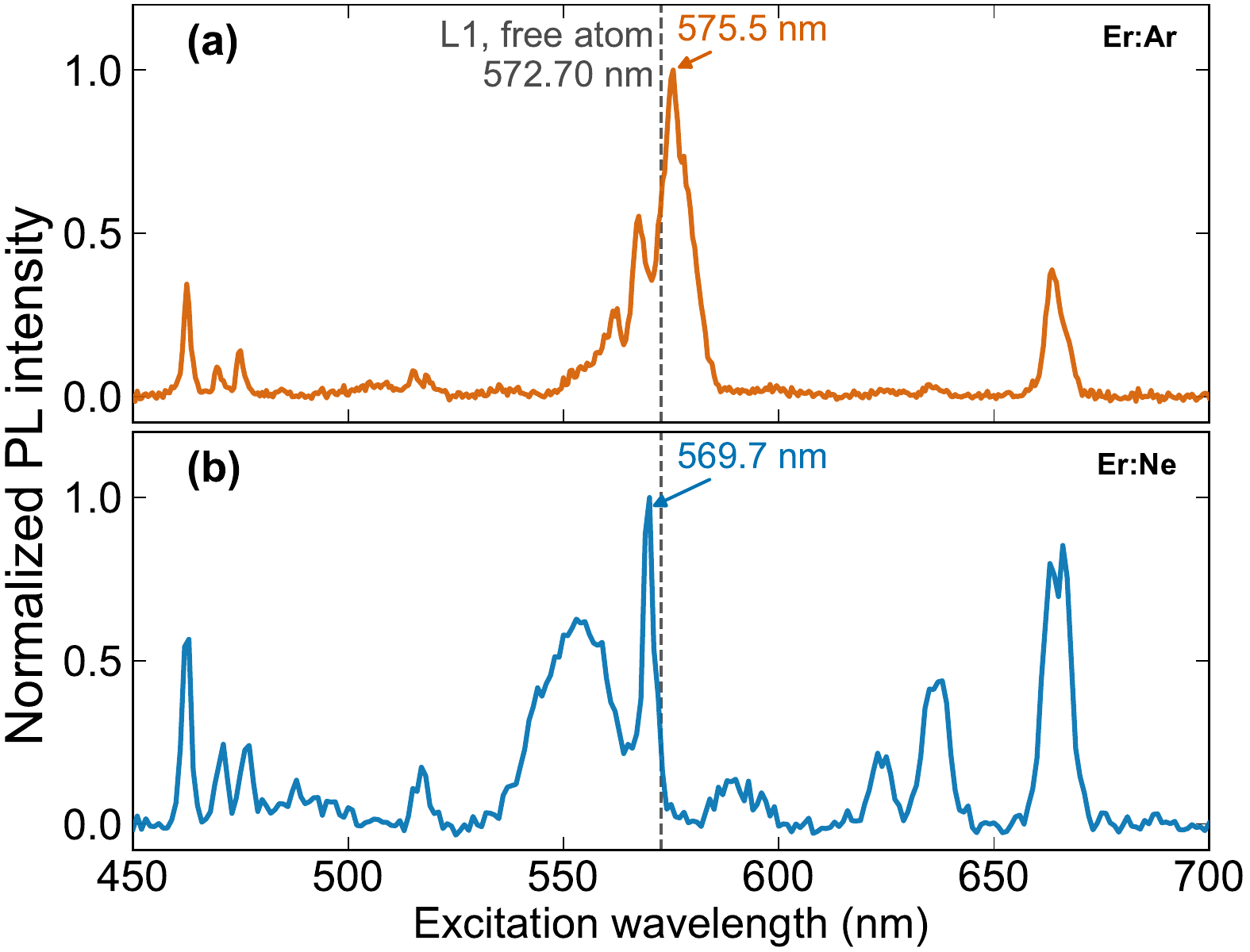}
    \caption{PLE spectra of Er in solid \textbf{(a)} Ar and \textbf{(b)} Ne, monitoring the L14 emission at \qty{1437}{\nm} before annealing. Each trace is background-subtracted and peak-normalized, without correction for the wavelength dependence of the excitation power. The excitation bandwidth is about \qty{2}{\nm}. The dashed vertical line marks the free-atom wavelength of L1 (\qty{572.70}{\nm}). Arrows mark the excitation wavelengths used for the PL decay measurements, \qty{575.5}{\nm} in Ar and \qty{569.7}{\nm} in Ne, near the maxima of the respective PLE spectra.}
    \label{Fig:Fig3}
\end{figure}

With \qty{200}{\um} slits, the measured L14 linewidth in Ne is about \qty{0.24}{\nm} (\qty{1.2}{\per\cm}). Narrowing the slits further reduces the measured width, so this value is limited by the instrumental resolution and represents an upper bound on the ensemble linewidth. Higher-resolution spectra give a narrower width of about \qty{0.15}{\nm} (\qty{22}{\GHz}) after annealing (Sec.~\ref{Sec:Annealing}). They also reveal at least two partially overlapping components, L14a and L14b, in both matrices (discussed later). Their separation depends on the host, and their relative weights change with annealing. Possible origins include crystal-field splitting and inequivalent trapping sites, which the present spectra cannot distinguish.

The PLE spectra of L14 (Fig.~\ref{Fig:Fig3}) show a maximum at \qty{575.5}{\nm} in Ar and \qty{569.7}{\nm} in Ne. Both lie close to the upper level of L1, the $(13/2,5/2)^{\circ}_{6}$ level at \qty{17456.383}{\per\cm}, whose free-atom air wavelength is \qty{572.697}{\nm}~\cite{atomic_energy_levels}. Relative to this level, the maximum is shifted by about $+\qty{92}{\per\cm}$ in Ne and $-\qty{85}{\per\cm}$ in Ar, which suggests that the enhancement reflects matrix-shifted absorption on this atomic transition. Although L1 emission itself is weak, this level efficiently feeds ${}^3\mathrm{H}_5$ through the relaxation pathways discussed above.

An emission feature also appears near \qtyrange{1187}{1190}{\nm} (Fig.~\ref{Fig:Fig2}(c,g)). Two E1-allowed transitions are candidates: from the odd-parity levels at \qty{24883.530}{\per\cm} ($J=8$) and \qty{25159.143}{\per\cm} ($J=7$) to the even-parity levels at \qty{16464.934}{\per\cm} ($J=7$) and \qty{16727.479}{\per\cm} ($J=8$), with free-atom air wavelengths of \qty{1187.521}{\nm} and \qty{1185.681}{\nm}, respectively~\cite{atomic_energy_levels}. Relative to the central Ne feature at \qty{1187.1}{\nm}, these correspond to matrix shifts of about $+\qty{3.0}{\per\cm}$ and $-\qty{10.1}{\per\cm}$. Both upper levels lie above the pumped level, so neither can be populated after single-photon absorption at \qty{532.2}{\nm}. The feature appears under strong \qty{532.2}{\nm} excitation but not under weaker broadband excitation, which suggests a two-step process via the long-lived ${}^3\mathrm{H}_5$ level. A second \qty{532.2}{\nm} photon absorbed from ${}^3\mathrm{H}_5$ reaches about \qty{25740}{\per\cm}, above both candidate upper levels. L9 shows a similar intensity dependence, suggesting that such two-step processes contribute to the population of several levels.

\begin{figure*}
    \centering
    \includegraphics[width=\textwidth]{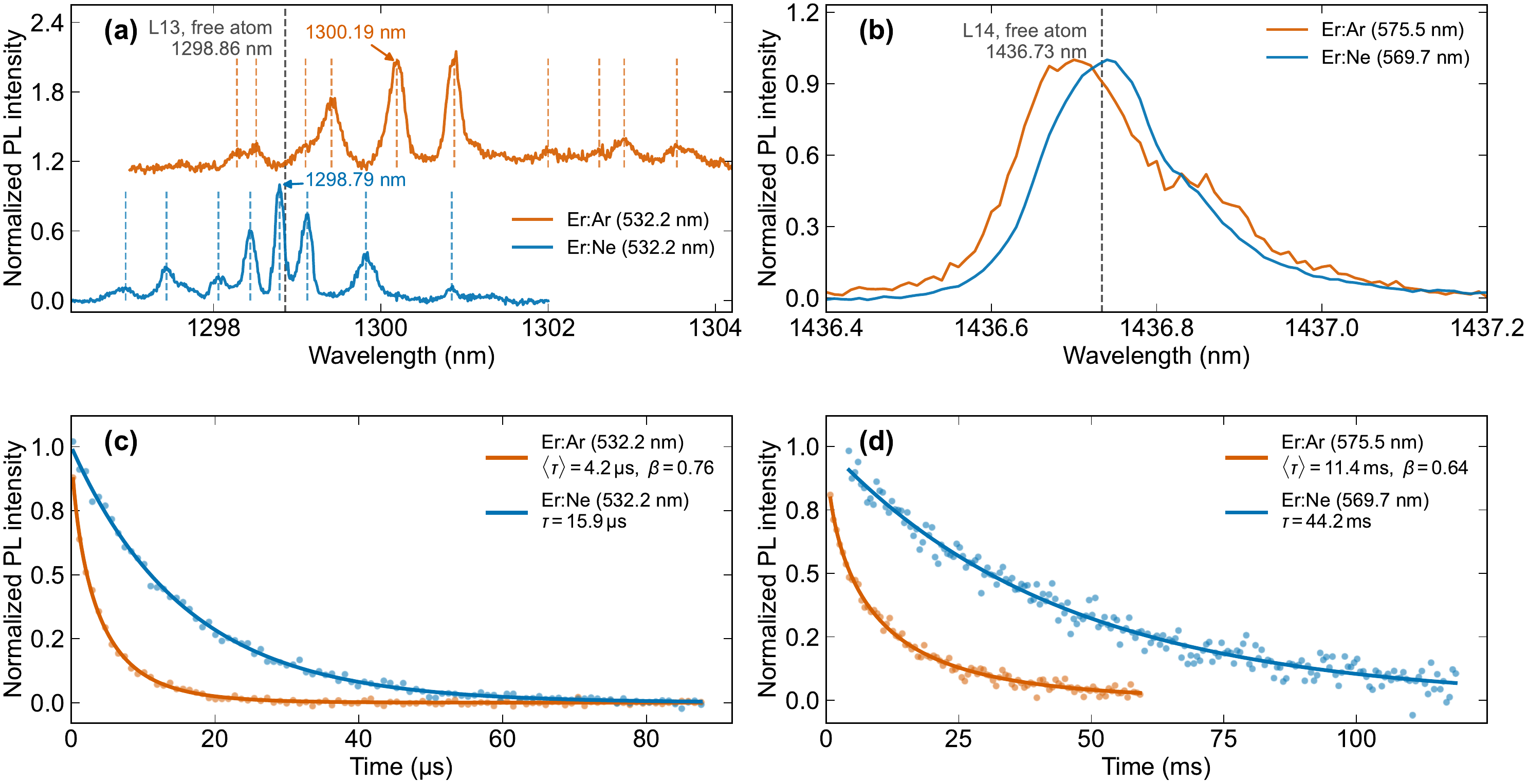}
    \caption{Spectra and decay dynamics of the telecom transitions L13 and L14 in as-grown Er:Ar and Er:Ne samples. \textbf{(a)} PL spectra of L13 near \qty{1299}{\nm} in solid Ar and Ne under \qty{532.2}{\nm} excitation. The Ar trace is vertically offset for clarity. Colored dashed lines mark the features listed in Table~\ref{tab:1299_components}, and arrows indicate the detection wavelengths used in (c). \textbf{(b)} PL spectra of L14 near \qty{1437}{\nm} under excitation at \qty{575.5}{\nm} (Ar) and \qty{569.7}{\nm} (Ne). In (a) and (b), spectra are baseline-corrected and peak-normalized, and gray dashed lines mark the free-atom transition wavelengths in air. \textbf{(c, d)} PL decays of L13 and L14, respectively, with time referenced to the end of the excitation gate. Solid lines are fits with a stretched exponential for Ar and a single exponential for Ne. The legends give the mean decay time $\langle\tau\rangle$ and stretching exponent $\beta$ for Ar, and the decay time $\tau$ for Ne, together with the excitation wavelengths in parentheses.}
    \label{Fig:Fig4}
\end{figure*}

L13 and L14 show strongly contrasting behavior (Fig.~\ref{Fig:Fig4}). L13 displays a host-dependent multiplet and microsecond-scale decay, whereas L14 remains close to its free-atom wavelength and decays on a millisecond scale. These two transitions therefore provide the clearest comparison between a predominantly $5d$-derived transition and a shielded $4f \rightarrow4f $ transition in the two matrices. Within the investigated spectral range and excitation conditions, we found no PL features that could be confidently assigned to Er ions.

\subsection{Lifetime measurement}
\label{Sec:Lifetime}

Time-resolved PL measurements reveal a pronounced contrast between the microsecond-scale decay of L13 and the millisecond-scale decay of the shielded $4f^{12}6s^{2}\,{}^{3}\mathrm{H}_{5}\rightarrow{}^{3}\mathrm{H}_{6}$ transition L14. PL photons were recorded by single-photon timestamping after gated excitation. For L13, we used \qty{532.2}{\nm} excitation and detected at \qty{1300.19}{\nm} in Ar and \qty{1298.79}{\nm} in Ne, as indicated in Fig.~\ref{Fig:Fig4}(a). For L14, the supercontinuum source provided excitation at the PLE maxima, \qty{575.5}{\nm} in Ar and \qty{569.7}{\nm} in Ne (Fig.~\ref{Fig:Fig3}). Excitation of L14 at \qty{532.2}{\nm} gave no noticeable change in the decay times. The supercontinuum source is pulsed, but its pulse spacing is much shorter than the PL decay time. The excitation is therefore effectively continuous within the AOTF gate, which defines the excitation envelope.

The decay curves in Fig.~\ref{Fig:Fig4}(c, d) were fitted with a free constant background $B$. For each host, we chose the simplest model that reproduces the data: single-exponential fits describe the Ne traces well, whereas the Ar traces require a stretched exponential. The Ne traces were fitted with
\begin{equation}
    I(t)=B+A\exp(-t/\tau).
\end{equation}
For the Ar traces, we used the Kohlrausch--Williams--Watts (KWW) stretched exponential,
\begin{equation}
    I(t)=B+A\exp\left[-\left(\frac{t}{\tau_{\mathrm{KWW}}}\right)^{\beta}\right],
\end{equation}
where $\tau_{\mathrm{KWW}}$ is the characteristic decay time and $\beta=1$ recovers a single exponential. The parameter $\beta$ describes non-exponential decay without assuming a specific number of emitting populations. For $0<\beta<1$, the KWW function corresponds to a continuous distribution of exponential decay rates~\cite{BerberanSantos2005Kohlrausch,Johnston2006Stretched}, which is physically plausible for matrix-isolated atoms in different local environments. However, a KWW fit alone cannot distinguish such heterogeneity from unresolved relaxation or feeding pathways. For the stretched-exponential fits, we report the effective mean decay time, defined as the time integral of the normalized fitted decay,
\begin{equation}
    \langle\tau\rangle
    =\int_{0}^{\infty}\exp\left[-\left(\frac{t}{\tau_{\mathrm{KWW}}}\right)^{\beta}\right]\mathrm{d}t
    =\frac{\tau_{\mathrm{KWW}}}{\beta}\,\Gamma\!\left(\frac{1}{\beta}\right),
\end{equation}
where $\Gamma$ is the gamma function.

For L13, the Ar fit gives $\langle\tau\rangle=\qty{4.2}{\us}$ with $\beta=0.76$, whereas the Ne fit gives $\tau=\qty{15.9}{\us}$ (Fig.~\ref{Fig:Fig4}(c)). These decays were measured only on the strongest components, Ar5 and Ne5 in Table~\ref{tab:1299_components}. For L14, the Ar fit gives $\langle\tau\rangle=\qty{11.4}{\ms}$ with $\beta=0.64$, and the Ne fit gives $\tau=\qty{44.2}{\ms}$ (Fig.~\ref{Fig:Fig4}(d)).

Both transitions decay more slowly in Ne than in Ar, by a similar factor of about 4. In both hosts, L14 is longer-lived than L13 by roughly three orders of magnitude. Together with their contrasting spectral structure, this clearly distinguishes the predominantly $5d$-derived and the shielded $4f$ emission channels. For L13, the matrix lifetimes are about four orders of magnitude shorter than the free-atom value of \qty{178}{\ms}, indicating strong nonradiative quenching. This is consistent with the \qty{520}{\per\cm} gap below its upper level, which requires fewer phonons to bridge in Ar than in Ne. The decay times alone, however, do not fully separate radiative decay, nonradiative relaxation, population transfer, and contributions from different trapping sites.

\subsection{Annealing and temperature-dependent PL}
\label{Sec:Annealing}

We further used thermal cycling to study how matrix preparation and local structural relaxation affect the PL of isolated Er atoms. In the rapid annealing procedure, the samples were heated from \qty{6}{\K} to nominally \qty{15}{\K} for Ne or \qty{40}{\K} for Ar, close to the sublimation regime under our conditions, and then rapidly cooled back to \qty{6}{\K}. Among the procedures we tested, these brief thermal excursions produced the clearest spectral changes. Most annealing procedures reduced the overall Er-related PL signal, but individual spectral features could still be enhanced, as for L13 in Ar under the conditions shown here. All spectra before and after annealing were recorded at \qty{6}{\K}. This separates persistent changes caused by thermal cycling from the immediate effects of a higher measurement temperature.

After annealing, both matrices showed spatially inhomogeneous optical properties. PL decay times measured at different positions on the same sample varied by up to about \qty{30}{\percent}, whereas no comparable variation was observed before annealing. This is consistent with annealing producing locally different emitter environments, although it does not identify their microscopic origin. This variation between sample positions does not contradict the narrowing of the L14 line in Ne discussed below, which reflects the distribution of environments within a single probed region. The variation also makes careful spatial registration important when comparing measurements before and after annealing, since a change in the probed region can alter the apparent spectral shape, intensity, and decay time.

\begin{figure*}
    \centering
    \includegraphics[width=\textwidth]{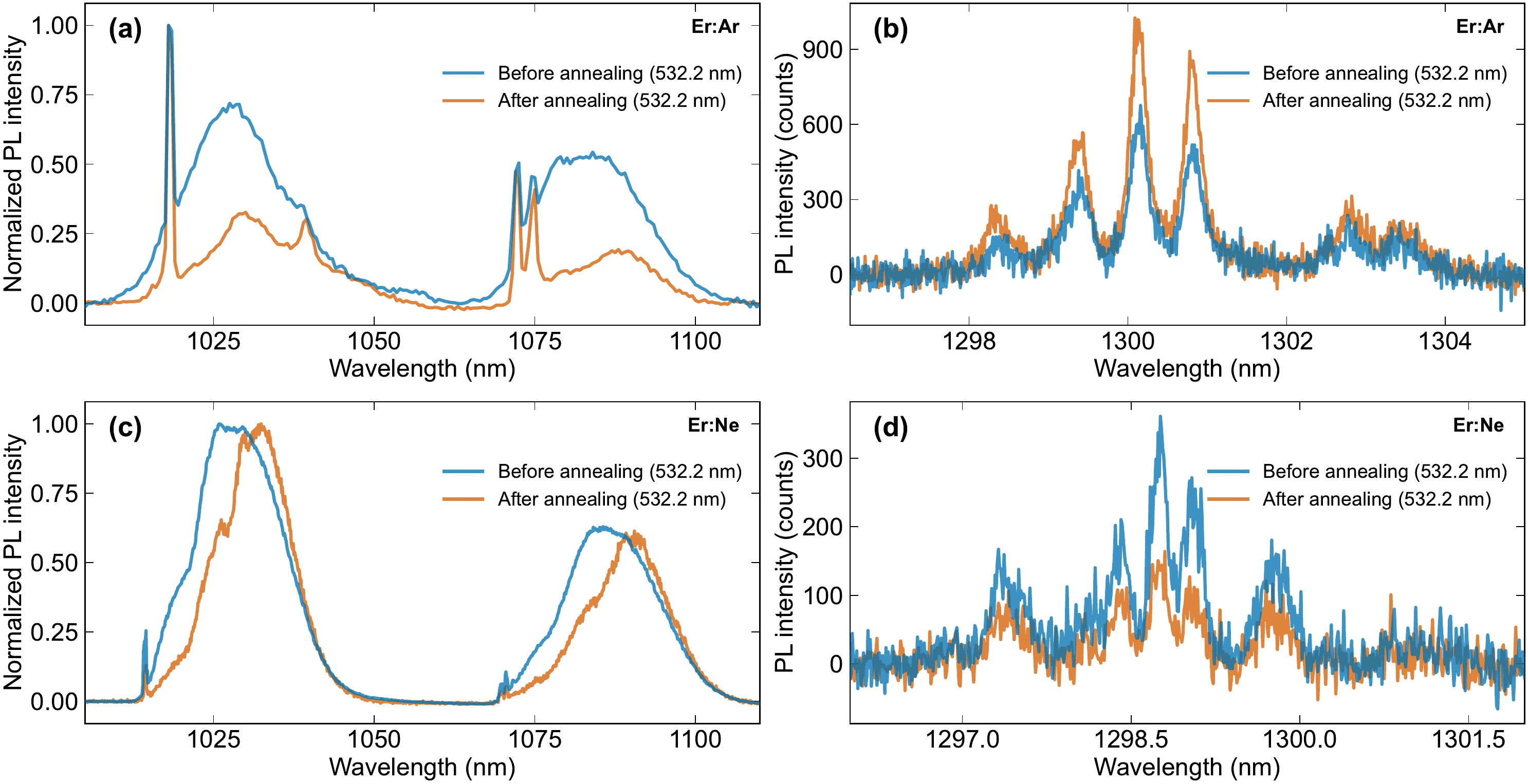}
    \caption{Effect of annealing on the PL spectra of Er in solid Ar \textbf{(a, b)} and Ne \textbf{(c, d)} under \qty{532.2}{\nm} excitation. Blue and orange traces show measurements before and after annealing, respectively. \textbf{(a, c)} NIR spectra around L11 and L12, baseline-corrected and peak-normalized to compare the band shapes and the relative prominence of the narrow features. \textbf{(b, d)} Detailed spectra of L13, shown as baseline-subtracted counts without peak normalization. Within each panel, the traces before and after annealing were recorded under the same acquisition conditions, but count scales should not be compared between hosts. Wavelengths in parentheses in the legends give the excitation wavelength.}
    \label{Fig:Fig5}
\end{figure*}

The two hosts respond differently in the NIR region around L11 and L12 (Fig.~\ref{Fig:Fig5}(a, c)). In Ar, annealing suppresses the broad bands, attributed to phonon sidebands (PSBs), relative to the narrow L11 and L12 zero-phonon lines (ZPLs). The ZPL/PSB peak-height ratio therefore increases, indicating a redistribution of spectral weight. This is consistent with changes in the local Er environments and in their coupling to the matrix.

For L13 in Ar, annealing increases the peak counts (Fig.~\ref{Fig:Fig5}(b)). Within the experimental resolution, the principal features neither shift nor change their relative intensities. Annealing therefore enhances the existing emission pattern rather than producing new components, which suggests a change in the excitation or emission efficiency of these features without a measurable change in their transition energies.

In Ne, annealing reduces the NIR emission by about an order of magnitude, shifts the maximum of the band near L11 from about \qty{1027}{\nm} to \qty{1032}{\nm}, and makes its substructure more pronounced (Fig.~\ref{Fig:Fig5}(c)). After annealing, local maxima appear at about \qty{1023}{\nm}, \qty{1026}{\nm}, and \qty{1029.5}{\nm}. Their spacing of about \qty{30}{\per\cm} is consistent with phonon replicas, although overlapping electronic or site-dependent contributions cannot be excluded. The ZPL/PSB peak-height ratio decreases by about \qty{40}{\percent}. In contrast, L13 retains essentially the same component positions and relative pattern (Fig.~\ref{Fig:Fig5}(d)). Although the count amplitudes differ, this spectral group shows no substantial reorganization.

\begin{figure*}
    \centering
    \includegraphics[width=\textwidth]{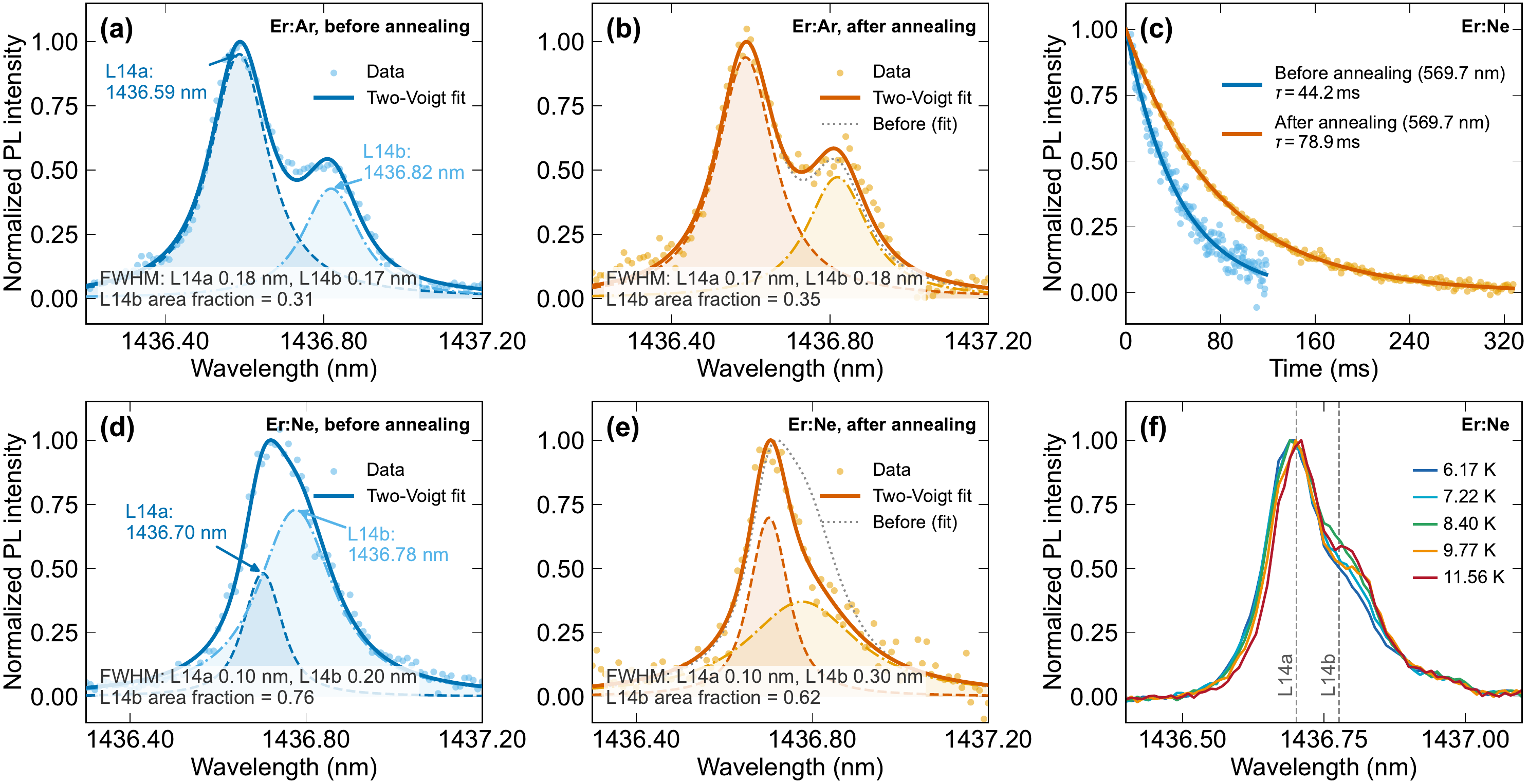}
    \caption{Annealing and temperature dependence of L14 at \qty{1437}{\nm}. \textbf{(a, b)} PL spectra of Er in solid Ar and \textbf{(d, e)} in solid Ne before and after annealing, under \qty{532.2}{\nm} excitation. Points show measurements, solid lines show two-Voigt fits, and dashed curves with shaded areas show the components L14a and L14b. For each host, the spectra before and after annealing were fitted simultaneously with shared component centers, labeled in (a) and (d), and free Lorentzian widths. The Gaussian width was fixed at \qty{0.05}{\nm}, so the fitted Lorentzian widths may include residual instrumental broadening. Annotations give the component widths (FWHM) and the area fraction $f_{\mathrm{L14b}}=A_{\mathrm{b}}/(A_{\mathrm{a}}+A_{\mathrm{b}})$. Dotted gray lines in (b) and (e) repeat the fit before annealing for comparison. Spectra are normalized to the fitted maximum. The decomposition is phenomenological, and the fitted widths are effective values. \textbf{(c)} Normalized Er:Ne PL decays before and after annealing under \qty{569.7}{\nm} excitation. Solid lines show single-exponential fits, with time referenced to the end of the excitation pulse. \textbf{(f)} Er:Ne spectra recorded during warming under \qty{569.7}{\nm} excitation, each normalized to its maximum. Dashed lines mark the L14a and L14b centers from (d, e).}
    \label{Fig:Fig6}
\end{figure*}

Annealing affects L14 differently in the two hosts (Fig.~\ref{Fig:Fig6}). To compare the spectra, we use a phenomenological two-component model and fit the spectra before and after annealing simultaneously, with shared component centers, a fixed Gaussian width, and free Lorentzian widths. The fitted parameters depend on the model constraints, so we compare the results qualitatively. In Ne, the emission intensity decreases by about an order of magnitude, while the line narrows from about \qty{0.20}{\nm} to \qty{0.15}{\nm} and its decay time increases from \qty{44.2}{\ms} to \qty{78.9}{\ms}. In Ar, the emission decreases to about half, and the line broadens from about \qty{0.33}{\nm} to \qty{0.37}{\nm}, with intensity moving from the line center into both wings. Within the model, the narrowing in Ne corresponds to a lower weight of the broad component L14b, which itself broadens. The broadening in Ar corresponds to a modest increase in the weight of L14b at nearly constant component widths. These different responses are consistent with the decay dynamics before annealing (Fig.~\ref{Fig:Fig4}(d)). The Ne decay is single-exponential, whereas the Ar decay is stretched and about four times faster, indicating a broader distribution of environments with stronger nonradiative quenching. Annealing thus appears to produce more uniform and less quenched environments within the probed region in Ne, but not in Ar.

The temperature series in Fig.~\ref{Fig:Fig6}(f) shows how the Ne L14 emission responds during warming, rather than after a completed annealing cycle. Between \qty{6.17}{\K} and \qty{11.56}{\K}, the line maximum shifts monotonically to the red by about \qty{0.02}{\nm} (\qty{0.1}{\per\cm}), and the line shows no thermal broadening. The thermal shift of L14 is therefore small, as expected for a shielded $4f$ transition. The line shape also changes systematically: intensity moves from the blue flank of the main peak to the red side near L14b, while the far red tail remains unchanged. This redistribution contributes to the red shift of the line centroid. The relative weight of the red side increases by about \qty{20}{\percent}, and the integrated emission decreases by about \qty{18}{\percent}. Thermal population cannot explain this redistribution. If the two components arise from sublevels of the same upper level, their separation of about \qty{0.4}{\per\cm} is much smaller than $\kB T\approx\qtyrange{4}{8}{\per\cm}$. Their relative population would then change by only a few percent over this range, and warming would favor the higher-energy, blue component. The redistribution therefore suggests temperature-dependent changes in the local environment or in the relaxation of the emitters. These changes appear already during warming, before the annealing cycle is completed, indicating that structural relaxation in solid Ne begins at relatively low temperatures.

The different responses of the two hosts are consistent with differences in trapping geometry, defect populations, and the relaxation pathways available during thermal cycling. In Ne, the narrower line and longer lifetime after annealing, together with the spectral changes that appear already during warming, suggest that the lattice around the emitters relaxes toward more uniform environments. The strong decrease in Ne emission, by a similar factor for L14 and for the NIR bands, could reflect a loss of emitters through diffusion or aggregation during annealing. The gradual decrease observed during warming suggests that this loss may begin before the annealing cycle is completed. In Ar, the broader line after annealing and the stretched, faster decay before annealing indicate that the emitters occupy a wider distribution of environments. The microscopic nature of the trapping sites remains open, since the size and rigidity of an Er trapping cavity do not follow simply from the size of the host atoms. Studies of other metal impurities show that Ne can host multi-vacancy trapping sites, whose geometry depends on the impurity--host interaction as well as on the host lattice~\cite{LaraMoreno2019TrappingSites}.

\subsection{Discussion}

In matrix-isolated Er, emission dynamics are driven more by the energy gaps between levels than by free-atom line strengths. Multiphonon relaxation in Ne and Ar requires many phonons to bridge gaps of more than a few hundred \unit{\per\cm}, so population collects on levels that lie above large gaps. This explains the dominant visible lines, the strong quenching of L13, and the absence of emission from the $J=6$ level at \qty{7177}{\per\cm}. It also identifies the ${}^3\mathrm{H}_5$ level, which lies \qty{1923}{\per\cm} above ${}^3\mathrm{F}_4$, as the principal bottleneck in the lower manifold and the source of the long-lived L14 emission.

To assess whether L14 approaches the radiative limit, we compare it with a closely related free-atom transition. L14 is a magnetic-dipole M1 transition between the fine-structure levels of the ${}^3\mathrm{H}$ term, analogous to the ${}^2\mathrm{F}_{5/2} \rightarrow {}^2\mathrm{F}_{7/2}$ transition in neutral Tm at \qty{1140}{\nm}, which has a free-atom lifetime of \qty{112}{\ms}~\cite{PhysRevA.94.022512}. This analogy suggests a free-atom radiative lifetime on the order of \qty{100}{\ms} for the transition underlying L14. In a host matrix with refractive index $n$, the M1 radiative decay rate scales as $n^3$~\cite{PhysRevB.83.245123}, shortening the lifetime only moderately in Ne but by approximately a factor of two in Ar~\cite{PhysRevB.11.1705, PhysRev.181.1297}.

The measured decay time of \qty{79}{\ms} in annealed Ne (Fig.~\ref{Fig:Fig6}(c)) is consistent with this estimate, suggesting predominantly radiative decay after annealing. By contrast, the substantially shorter decay times in as-grown Ne and Ar indicate significant nonradiative losses.

L13 exhibits the opposite behavior: although its free-atom lifetime of \qty{178}{\ms} is comparable to that expected for L14, its matrix lifetimes are shorter by approximately four orders of magnitude. This reduction is attributed to nonradiative relaxation across the small energy gap below its upper level, rather than to a comparable increase in the radiative decay rate.

The two telecom transitions also differ markedly in their sensitivity to the crystal field. L13 splits into a multiplet spanning approximately \qty{23}{\per\cm} in Ne and \qty{31}{\per\cm} in Ar, whereas the L14 components are separated by only about \qty{1}{\per\cm} or less. Because both transitions terminate on the ${}^3\mathrm{H}_6$ ground level, the substantially larger splitting of L13 must originate predominantly from its upper level. This level contains a $5d$ electron whose spatially extended orbital penetrates beyond the shielding $5s^2$ and $5p^6$ subshells and is therefore much more strongly coupled to the surrounding host atoms.

If the observed L13 multiplet reflects crystal-field splitting of this $J=7$ level rather than occupation of inequivalent trapping sites, the splitting exceeds that of the $4f^{12}$ levels involved in L14 by more than an order of magnitude. This contrast provides a direct indication of the much stronger host interaction experienced by the $5d$ electron relative to the shielded $4f$ electrons. Even for the $5d$ electron, however, the splitting remains far smaller than in ionic crystals, where $5d$ crystal-field splittings of rare-earth ions typically reach thousands of \unit{\per\cm}~\cite{PhysRevB.65.045114}, highlighting the weak perturbation provided by rare-gas hosts.

The two hosts therefore offer contrasting environments. In Ne, the weak host perturbation, the near-radiative L14 decay after annealing, and the narrowing of its line indicate that the emitters can occupy uniform, weakly quenched environments. In Ar, the broader lines and stretched decays point to a wider distribution of environments and stronger coupling to the lattice. The two-component structure of L14 in both hosts remains open: it may reflect inequivalent trapping sites or crystal-field splitting, and the similar excitation spectra of the two components slightly favor the latter.

\subsection{Conclusion}

We have isolated neutral Er atoms in solid Ne and Ar and characterized their optical emission from the visible to the telecom range. The observed spectra are explained by the free-atom level structure combined with multiphonon relaxation, which channels population into levels above large energy gaps. The shielded $4f\rightarrow4f$ transition L14 at \qty{1437}{\nm}, in the telecom E band, shows a negligible matrix shift, a narrow line, and a lifetime of up to \qty{79}{\ms} in annealed Ne, close to its estimated radiative limit. The $5d$-derived transition L13 at \qty{1299}{\nm}, in the O band, shows much stronger host coupling, with a crystal-field multiplet and microsecond-scale decay. Annealing affects the two hosts differently, producing more uniform, weakly quenched environments in Ne but not in Ar. Combined with the nearly nuclear-spin-free environment of rare-gas solids, these properties make Er in solid Ne a promising platform for long-lived telecom emitters.

\ \\
\iffalse

\noindent\textbf{Data availability}

The data that support the findings of this study are available from the corresponding authors upon request. Source data are provided with this paper.\\

\noindent\textbf{Code availability}

The codes used to perform the experiments and to analyze the data in this work are available from the corresponding authors upon request.\\

\fi

\acknowledgements

E. Z., Y. W., and D. J. acknowledge support from the Industry-University Cooperative Research Center Program at the U.S. National Science Foundation (NSF) under Grant No. 2224985. E. Z. and D. J. acknowledge support from the NSF under Award No. DMR-2415737. S. Z., A. D., Y. W., and D. J. acknowledge support from the Air Force Office of Scientific Research (AFOSR) under Award No. FA9550-23-1-0636. C. P., W. G., and D. J. acknowledge support from the NSF under Award No. OSI-2426768. C. P. and W. G. acknowledge the support from the National High Magnetic Field Laboratory at Florida State University, which is supported by National Science Foundation Cooperative Agreement No. DMR-2128556 and the State of Florida. Work performed at the Center for Nanoscale Materials, a U.S. Department of Energy Office (DOE) of Science User Facility, was supported by the U.S. DOE, Office of Basic Energy Sciences, under contract No. DEAC02-06CH11357.\\

\newpage
\bibliography{references}

\end{document}